\documentclass[aps,prl,twocolumn,superscriptaddress]{revtex4-1}

\usepackage{graphicx}
\usepackage{dcolumn}
\usepackage{bm}

\begin{document}
	\newcommand{\kp}{{\bf k$\cdot$p}\ }
	\title{Controlled electron transmission by lead
		chalcogenide barrier potential}

	\author{P. Pfeffer}\email{pfeff@ifpan.edu.pl} \affiliation{Institute of Physics, Polish Academy of Sciences, Aleja Lotnikow 32/46, PL-02668 Warsaw, Poland}
	\author{W. Zawadzki}\affiliation{Institute of Physics, Polish Academy of Sciences, Aleja Lotnikow 32/46, PL-02668 Warsaw, Poland}
	\author{K. Dybko}\affiliation{Institute of Physics, Polish Academy of Sciences, Aleja Lotnikow 32/46, PL-02668 Warsaw, Poland}	
	\affiliation{International Research Centre MagTop, Institute of Physics, Polish Academy
		of Sciences, Aleja Lotnikow 32/46, PL-02668 Warsaw, Poland}
	
	\begin{abstract}
	 Transmission of electrons across a rectangular barrier of IV-VI semiconductor compounds is considered.  Conduction electrons arrive at the barrier and are reflected or transmitted through it depending on the relative values of the barrier potential $V_b$ and the electron energy $E$. The theory, in close analogy to the Dirac four component spinors, accounts for the boundary conditions on both sides of the barrier. The calculated transmission coefficient $T_C$ is an oscillatory function of the barrier voltage varying between zero (for full electron reflection) and unity (for full electron transmission). Character of electron wave functions outside and inside the barrier is studied. There exists a total current conservation, i. e. the sum of transmitted and reflected currents is equal to the incoming current. The transmission  $T_C$ is studied for various barrier widths and incoming electron energies. Finally, the transmission coefficient $T_C$ is studied as a function of $V_b$ for decreasing energy gaps $E_g$  of different Pb$_{1-x}$Sn$_x$Se compounds in the range of 150 meV $\geq E_g \geq$ 2 meV. It is indicated that for very small gap values the behaviour of $T_C$ closely resembles that of the chiral electron tunneling by a barrier in monolayer graphene. For $E_g$ =0 (Pb$_{0.81}$Sn$_{0.19}$Se) the coefficient $T_C$ reaches the value of 1 independently of $V_b$.
	\end{abstract}  

	\maketitle
	
	\section{Introduction}
Narrow gap IV-VI compounds PbS, PbSe and PbTe are among the oldest known semiconductor compounds used for electricity and electronics. In particular, PbS is probably the oldest known applicable semiconductor. These materials and their alloys with thin counterparts came recently into prominence as they constitute  topological crystalline insulators (TCI) material class, c.f. Dziawa et al \cite{PD}. Due to  heavy metal Pb the lead
chalcogenides possess strong spin-orbit interaction (SOI) and consequently are characterized  by large spin g-factors. Similarly to other narrow-gap materials (see e.g. III-V compounds), SOI results, in addition to {\kp} mixing between the valence and conduction bands, also in mixing of the spin states. This problem is discussed in the work of Ravich et al \cite{RA}.

In addition to the features mentioned above, the band structure of lead
chalcogenides at the L point of the Brillouin zone is almost spherical and strongly resembles that of the Dirac Hamiltonian \cite{Dirac1928} in that it consists of two symmetric conduction and valence bands. This similarity allows us to use for the description of these materials methods of relativistic quantum theory. 

It is known that semiconductors produce in general strong effects of the spin-orbit interaction. On the other hand, while in the Dirac Hamiltonian one deals with pure electron spins, in semiconductors one deals in similar situations with pseudospins. This relates to the spin-orbit interaction resulting from the periodic potential of the crystal lattice not present in vacuum.

In the present work we consider electron transmission controlled by PbS and  Pb$_{1-x}$Sn$_x$Se barrier potentials. Barrier penetration and transmission is an important phenomenon in many physical and chemical problems like interband and intraband tunneling behavior of mesoscopic systems, electronic transitions, nuclear decay, etc.

An interesting case of barrier penetration was proposed for monolayer and bilayer graphene by Katsnelson et al \cite{Kats2006} in which chiral and nonchiral electron tunnelling was considered using the Klein scattering through a rectangular barrier created by an electric potential. 
The theoretical aspects of Klein tunnelling in graphene were reviewed in Refs \cite{Allain2011,Das2011} and extended to other elemental two-dimensional materials \cite{phosphor2017, bor2018}.
Su et al \cite{Su93}, Calogeracos and Dombey \cite{Cal99,Dom99}  took into account relativistic effects in electron penetration of a one-dimensional barrier in vacuum.

	\section{THEORY}

Basic results on the band theory of IV-VI lead salts were obtained by Mitchell and Wallis \cite{Mitch}, Dimmock \cite{Dimm1971} and Grisar et al \cite{Grisar1978}. In these materials each of the four ellipsoids at the $L$ points of the Brillouin zone can be described by the {\kp} theory with the use of $X, Y$ and $Z$ functions $L^+_6$ providing secular equation for the energy  ${\cal E}(k)$:
\ \\
\begin{equation}
{\cal E}(E_g + {\cal E})= k_z^2 {P_{||}}^2 +(k_x^2 +k_y^2){P_{\perp}}^2\;\;,
\end{equation}
\ \\
where $P_{||}$ and $P_{\perp}$ are the interband  momentum matrix elements and $E_g$ is the forbidden gap between conduction and valenece bands. For PbS at low temperatures $E_g$ = 0.283 eV, the density of states mass  $m^*_d$ is $(m^*_{\parallel} {m^*_{\perp}}^2)^{1/3} = 0.0881 m_0$ and the conductivity effective mass  $m^*_c=3/({m^*_{\parallel}}^{-1}+2{m^*_{\perp}}^{-1})$ is $0.0873 m_0$. They are only slightly different, so we use the average conduction band mass $m^*_0=0.0877m_0$ corresponding to the matrix elements $P_{||} = P_{\perp} = P$. This effective mass is 
\ \\
\begin{equation}
\frac{m_0}{m^*_0} = \frac{m_0}{m_{CV}} + \frac{m_0}{m_{FB}}\;\;,
\end{equation}
\ \\
where the first term comes from the {\kp} interaction of the conduction and valence bands and $m_{FB}$ accounts for the contribution of far bands.
\ \\
The two-level anisotropic matrix Hamiltonian for carriers without far-band terms was given by Dimmock and Wright \cite{Dimm1964}.
The matrix in the isotropic approximation for $m^*_0$ is in general	
\ \\
\[\left[ \begin{array}{cccc}
\frac{E_g}{2}-\lambda &0&K_{||}&K_{\perp}\\
0&\frac{E_g}{2}-\lambda&K_{\perp}^*&-K_{||}\\
K_{||}&K_{\perp}&-\frac{E_g}{2}-\lambda &0 \\
K_{\perp}^*&-K_{||}&0&-\frac{E_g}{2}-\lambda
\end {array}\right]	\]
\ \\
where $K_{||}=(E_g\hbar^2/2m^*_0)^{1/2}k_z$, $K_{\perp}=(E_g\hbar^2/2m^*_0)^{1/2}(k_x+ik_y)$ and effective mass $m^*_0$ is

\begin{equation}
m^*_0 = \frac{3E_g \hbar^2}{4P^2}\;\;.
\end{equation} 
In our configuration $k_x = k_y = K_{\perp} =  K^*_{\perp} = 0$ 

\begin{figure}
	\includegraphics[width=1.\columnwidth, trim=-1.5cm 1.cm 0cm 0cm]{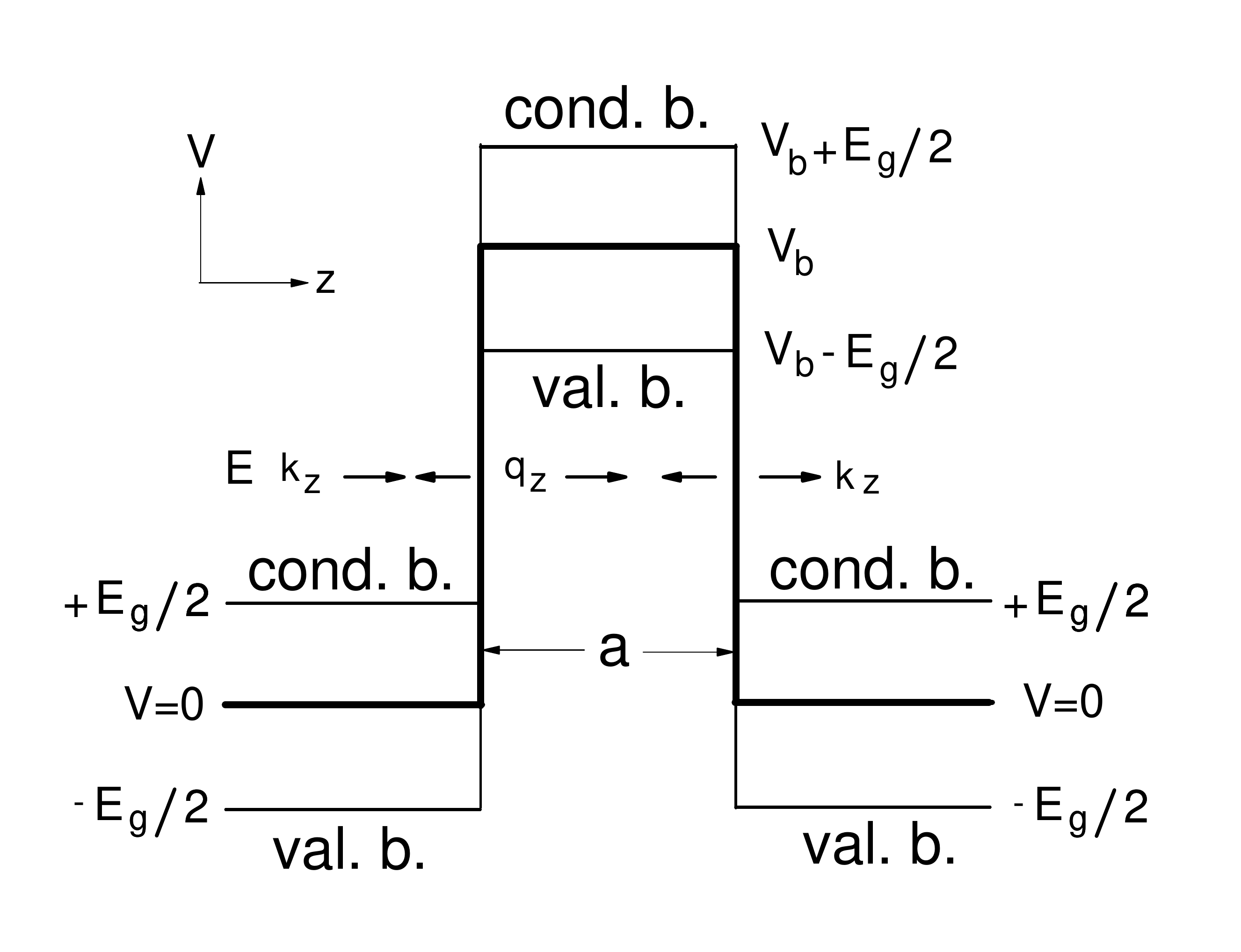}
	\caption{Rectangular potential illustrating the barrier scattering.
		Conduction electrons come from the left having the energy $E$ and momentum $\hbar k_z$. }
	\label{fig1th}
\end{figure}
\ \\
We consider conduction PbS electrons coming from the left along the $z$ direction to the rectangular barrier potential described by $V(z) = V_b$ for $0 < z < a$, where $a$ is the barrier width. Dependence of the conduction and valence bands and electron momentum on $V(z)$ are shown in Fig. 1.  For lead
chalcogenides the electron function contains two spinors for positive electron energies similarly to the Dirac equation \cite{WZ2017}.
The electron spin-up state (+) is

\begin{equation}
\Theta(+) = \frac{1}{|2({\cal E}-V) (E-V)|^{1/2}}
\left(\begin{array}{c}{\cal E}-V\\0\\K_z\\0\end{array}\right)
\;\;,
\end{equation}
\ \\
having the energy
\ \\
\begin{equation}
E =\left[\left(\frac{E_g}{2}\right)^2+E_g\left(\frac{\hbar^2 {k_z}^2}{2m^*_0}\right)\right]^{1/2}+V 
\;\;,
\end{equation}
\ \\
where ${\cal E}=E+E_g/2$ and $K_z=(E_g\hbar^2/2m^*_0)^{1/2}k_z$. 
\ \\
The spin-down state (-)
is 
\ \\

\begin{equation}
\Theta(-) = \frac{1}{|2({\cal E}-V) (E-V)|^{1/2}}
\left(\begin{array}{c}
0\\({\cal E}-V)\\0\\-K_z\end{array}\right)
\;\;,
\end{equation}
\ \\
and the energy $E$ is given by Eq.(5).

The momentum $\hbar k_z$ on the left and right of the barrier (for $V$ = 0) is
\begin{equation}
\hbar k_z = \left[[E^2-(\frac{E_g}{2})^2](\frac{2m^*_0}{E_g})\right]^{1/2}\;\;,
\end{equation}
\ \\
Inside the barrier (for $V = V_b$) the momentum $\hbar q_z$ is
\begin{equation}
\hbar q_z = \left[[(E-V_b)^2-(\frac{E_g}{2})^2](\frac{2m^*_0}{E_g})\right]^{1/2}\;\;.
\end{equation}

Suppose the electrons come to the barrier in the spin-up state (+) described by the function $\Psi(+)$. Since the electron reflection and transmission is elastic, there should exist a reflected wave on the left of the barrier, transmitted and reflected waves inside the barrier and a transmitted wave on the right of the barrier. Thus we have, see Eqs. (4) and (6)
\ \\
$$
\Psi(+) = N_k\left[ e^{ik_z z}
\left(\begin{array}{c}
{\cal E}\\0\\K_z\\0 
\end{array}\right)
+R_1 e^{-ik_z z}
\left(\begin{array}{c}
{\cal E}\\0\\-K_z\\0 
\end{array}\right)
\right]_{z < 0}+
$$
$$
+N_q\left[T_1 e^{iq_z z} 
\left(\begin{array}{c}
{\overline{\cal E}}\\0\\Q_z\\0 
\end{array}\right)
+R_2 e^{-iq_z z}
\left(\begin{array}{c}
{\overline{\cal E}}\\0\\-Q_z\\0 \end{array}\right)
\right]_{0\le z\le a}+
$$
\begin{equation}
+N_k
\left[T_2 e^{ik_z z} \left(\begin{array}{c}
{\cal E}\\0\\K_z\\0 
\end{array}\right)
\right]_{z > a}\;\;.
\end{equation}
\ \\
where we introduced the notation  $N_k = (2{\cal E}\cdot E)^{-1/2}$, $N_q = (2|{\overline{\cal E}}\cdot {\overline E}|)^{-1/2}$, $Q_z=(E_g\hbar^2/2m^*_0)^{1/2} q_z$,  ${\overline{\cal E}} = E- V_b+E_g/2$ and ${\overline E} = E-V_b$.

The coefficients $R_1$ and  $R_2$ are related to the reflected  waves, while $T_1$ and $T_2$ are those of the transmitted ones. These coefficients can be determined by the boundary conditions, i.e. by making equal each of the four spinor components in Eq. (9) at $z = 0$ and at $z = a$. One obtains

\begin{equation}
\frac{{\cal E}}{({\cal E}\cdot E)^{1/2}}(1+R_1)=\frac{\overline {\cal E}}{(|{\overline{\cal E}}\cdot {\overline E}|)^{1/2}}(T_1+R_2)\;,
\end{equation}
\ \\

\begin{equation}
\frac{K_z}{({\cal E}\cdot E)^{1/2}}(1-R_1)=\frac{Q_z}{(|{\overline{\cal E}}\cdot {\overline E}|)^{1/2}}(T_1-R_2)\;,
\end{equation}
\ \\
\begin{equation}
\frac{{\cal E}}{({\cal E}\cdot E)^{1/2}}T_2 e^{ik_z a} =\frac{\overline {\cal E}}{(|{\overline{\cal E}}{\overline E}|)^{1/2}}(T_1 e^{iq_z a}+R_2e^{-iq_z a})\;,
\end{equation}
\ \\
\begin{equation}
\frac{K_z T_2 e^{ik_z a}}{({\cal E}\cdot E)^{1/2}}(1-R_1)=\frac{Q_z}{(|{\overline{\cal E}}\cdot {\overline E}|)^{1/2}}(T_1 e^{iq_z a}-R_2 e^{-iq_z a})\;,
\end{equation}

In order to simplify subsequent formulas we introduce the so called kinematic factor $\kappa$. Employing Eqs. (7) and (8) one obtains
\begin{equation}
\kappa = \frac{Q_z{\cal E}}{K_z{\overline {\cal E}}} =
\frac{\{[E-V_b-E_g/2][E-V_b+E_g/2]\}^{1/2}} {\{[E-E_g/2][E+E_g/2]\}^{1/2}} \frac{{\cal E}}{\overline {\cal E}}\;\;.
\end{equation}
\begin{figure}
	\includegraphics[width=1.05\columnwidth, trim=-1.cm 1.cm 0.cm 0.cm]{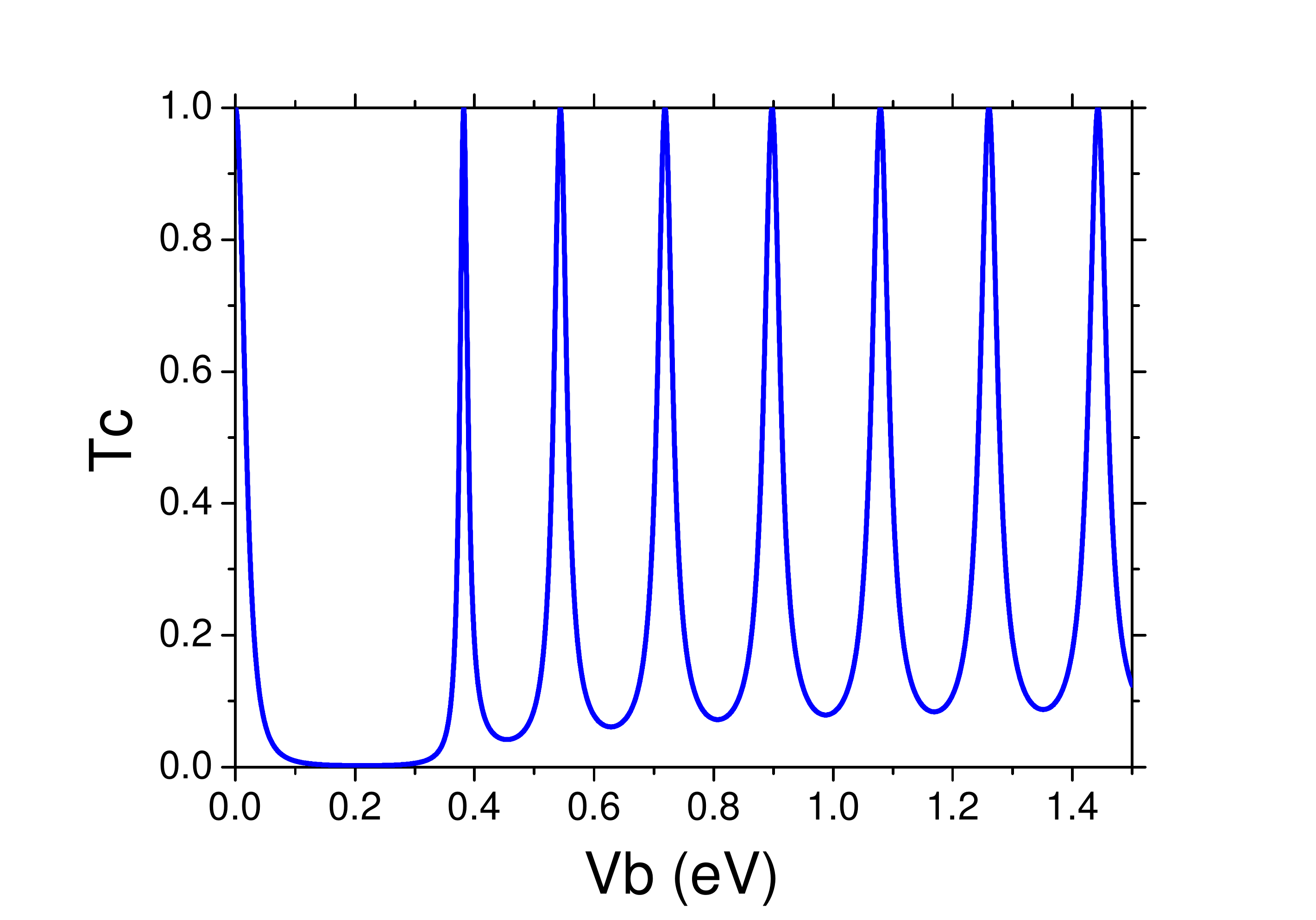}
	\caption{Oscillatory transmission probability $T_C$ versus applied voltage $V_b$ calculated for PbS barrier potential of the with $a = 6$ nm. Incoming electron energy is $E$ = 0.15 eV. For $ V_b $ from $(E - E_g/2)$ to $(E + E_g/2)$ the electron energy is within the forbidden gap of the barrier (see Fig. 1) and $ T_C $ is practically zero.}
	\label{fig2th}
\end{figure}
\ \\
With the use of $\kappa$ we finally have
\ \\
\begin{figure}
	\includegraphics[width=1.05\columnwidth, trim=-1.cm 1.cm 0cm 0cm]{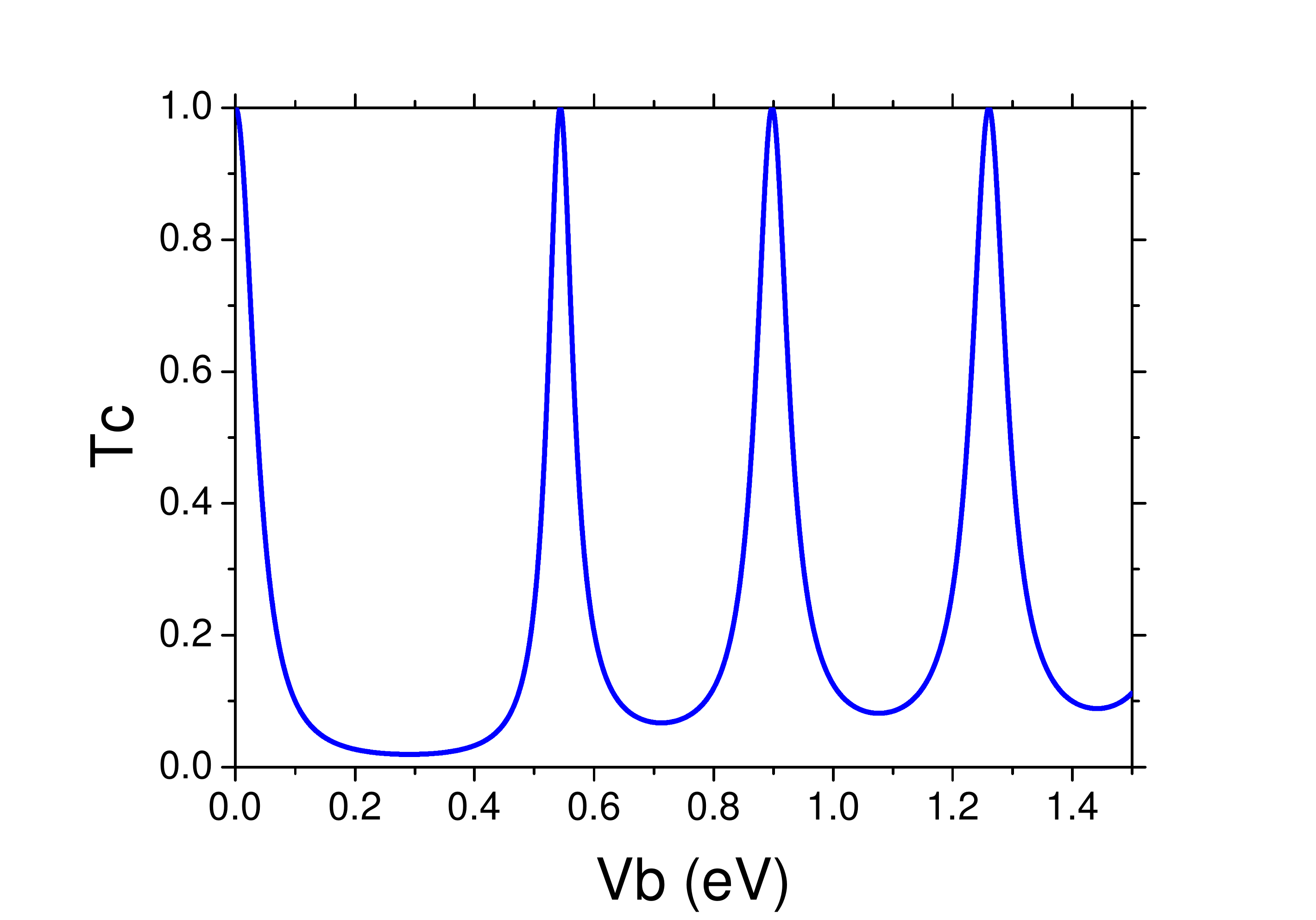}
	\caption{Oscillatory transmission probability $T_C$ calculated for PbS barrier of the width $a$ = 3 nm. For the narrower barrier the damping of transmission is weaker and the minimum of $T_C$ does not reach zero.}
	\label{fig3th}
\end{figure}
\begin{equation}
R_1=\frac{(1-\kappa^2)(1- e^{i2q_z a})}{[(1+\kappa)^2-(1-\kappa)^2 e^{i2q_z a}]}
\end{equation}
\ \\
\begin{equation}
T_1=\frac{2 A (1+\kappa)}{[(1+\kappa)^2-(1-\kappa)^2e^{i2q_z a}]}
\end{equation}
\ \\
\begin{equation}
R_2= \frac{-2A(1-\kappa)e^{i2q_z a}}{[(1+\kappa)^2-(1-\kappa)^2e^{i2q_z a}]}
\end{equation}
\ \\
\begin{equation}
T_2=\frac{4e^{-ik_z a}e^{iq_z a}\kappa}{[(1+\kappa)^2-(1-\kappa)^2 e^{i2q_z a}]}
\end{equation}
\ \\
\ \\
where $A=[|{\overline{\cal E}}\cdot {\overline E}|/({\cal E}\cdot E)]^{1/2}\cdot {\cal E}/{\overline{\cal E}}$.
 
The coefficients involved in $\Psi(+)$ and $\Psi(-)$ functions are the same.
\ \\
\subsection{CHARACTER OF WAVE FUNCTIONS}
\ \\

Next we consider the character of wave functions. This character is determined by relative values of $V_b$ and the electron energy $E$, see Fig. 1. The components outside the barrier have the plain wave character. However, the components inside the barrier can have either the plain wave or decaying character. This depends  on the momentum $\hbar q_z$ in the $V_b$ region since $\hbar q_z$, given by Eq. (8), can be real or imaginary. In both spin states, (+) and (-) $q_z$ is given by the relation
\ \\
\begin{figure}
	\includegraphics[width=1.05\columnwidth, trim=-1.cm 1.cm 0.cm 0.cm]{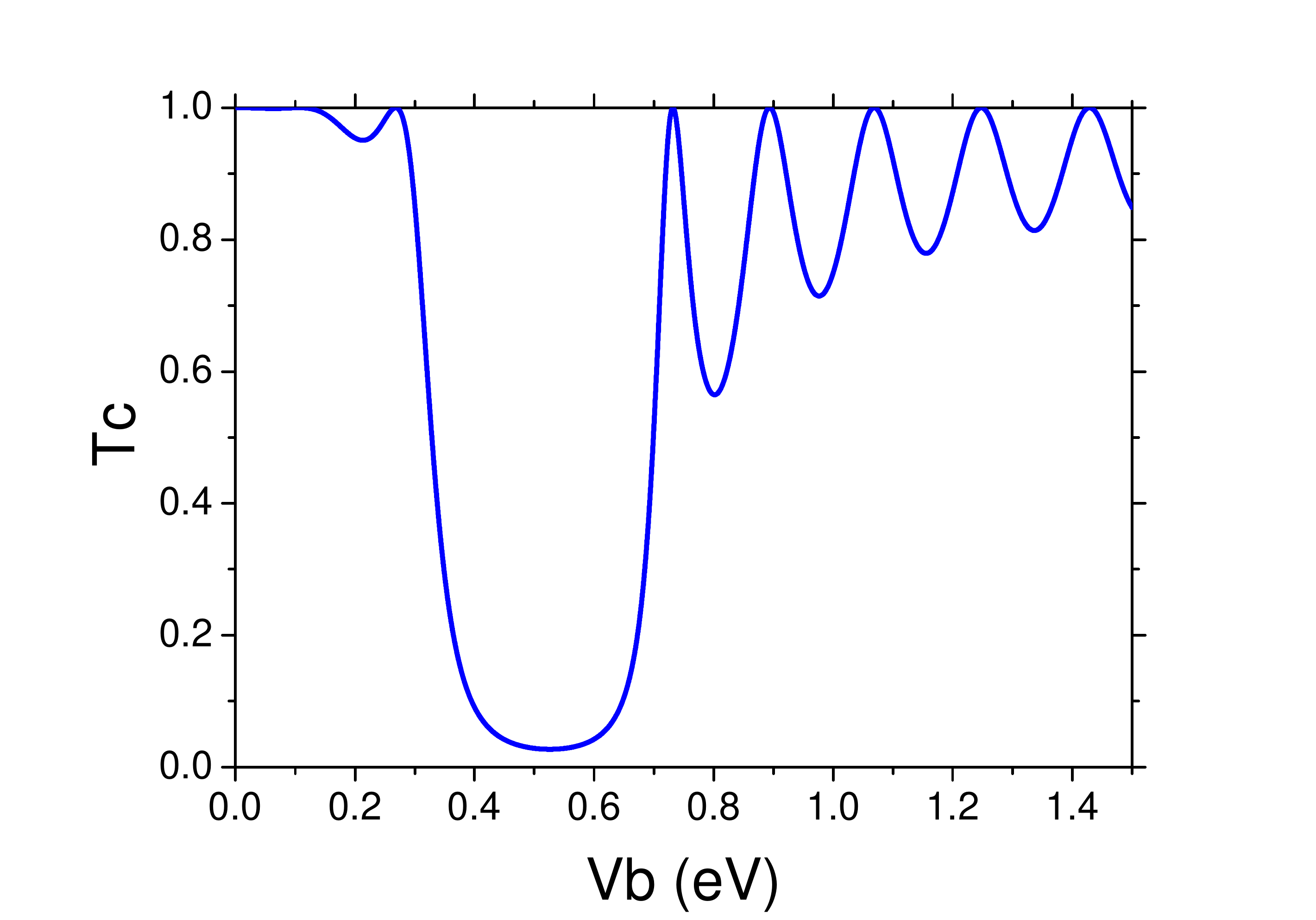}
	\caption{Oscillatory transmission probability $T_C$ calculated for electron energy $E$ = 0.5 eV and PbS barrier width $a$ = 6 nm. A difference in the dependences of $T_C$ on $V_b$ in Figs 2 and 4 result from  differences in the value of $E-E_g/2$ and consequently in the values of $\kappa$, (see Eq. (14)).}
	\label{fig4th}
\end{figure}
\ \\
\begin{equation}
\frac{E_g\hbar^2 {q_z}^2}{2m^*_0} = (E - V_b - \frac{E_g}{2})(E - V_b + \frac{E_g}{2})\;\;.
\end{equation}
\ \\

One can define two important cases for the momentum $\hbar q_z$:
\ \\

Case I. $ V_b - \frac{E_g}{2} > E > V_b + \frac{E_g}{2}$, so that $q_z$ and $\kappa$ are real numbers, see Eqs (14) and (19). In consequence, the transmitted amplitude $T_2$ in Eq. (17) is the plain wave and the transmission coefficient $T_C=|T_2|^2$ is 
\ \\
\begin{equation}
T_C=\frac{4\kappa^2}{[4\kappa^2+(1-\kappa^2)^2sin^2(q_z a)]}\;\;,
\end{equation}
\ \\
while reflection coefficient $R_C=|R_1|^2$ is
\ \\
\begin{equation}
R_C=\frac{(1-\kappa^2)^2sin^2(q_z a)}{[4\kappa^2+(1-\kappa^2)^2sin^2(q_z a)]}\;\;,
\end{equation}
so that $T_C + R_C =1$. The coefficients $T_C$ and $R_C$ are periodic functions of $q_z$ and for $sin^2( q_z a) = 0$ $T_C=1$ and $R_C=0$, while for $sin^2( q_z a)=1$ $T_C$ reaches the minimum value
\begin{equation}
T_C^{min} = 4\kappa^2/(1+\kappa^2)^2\;\;,
\end{equation}
and $R_C$ reaches the maximum value = $(1-\kappa^2)^2/(1+\kappa^2)^2$. The fact that the electron passes the barrier without any reflection for certain $q_z$ values is possible because the phase of the electron plain wave $e^{i q_z z}$ for $z = a$ is N $\pi$  where N = 0, $\pm$1, $\pm$2... . A comparison of Figs 2 and 3 shows that $T_C$ vanishes more quickly with increasing value of $a$.
\ \\
\subsection{CURRENT CONSERVATION}
\ \\
The sum of reflected and transmitted currents should be equal to the incident current:
\begin{equation}
j_{inc}=j_{R_C}+j_{T_C}\;\;,
\end{equation}
where each term is given by the electron charge multiplied  by the probability density and the  group velocity $v_{gr} = \partial E/\partial \hbar k_z = E_g\hbar k_z/2m^*_0 E$. Using Eq. (23) one has
\begin{equation}
\frac{eE_g\hbar k_z}{2m^*_0 E}=\frac{eE_g\hbar k_zR_C}{2m^*_0 E}+\frac{eE_g\hbar k_zT_C}{2m^*_0 E}
\end{equation}
which is equivalent to

\begin{equation}
1=R_C+T_C \;\;.
\end{equation}
Thus the current conservation is fulfilled.

\ \\
Case II. $ V_B + \frac{E_g}{2} > E > V_B - \frac{E_g}{2}$. Now $q$ and $\kappa$ are imaginary numbers. In consequence, the transmission amplitude $T_2$ and the transmission coefficient $T_C$ are decaying functions depending on the value of $a$ (see Figs 2 and 3). The coefficient $T_C$ is (see Eq.(18))
\ \\
\begin{equation}
T_C=\frac{16\kappa^2e^{-2|q_z| a}}{[(1+|\kappa|^2)^2(1-e^{-2|q_z| a})^2+16|\kappa|^2e^{-2|q_z| a}]}\;\;,
\end{equation}
\ \\
while the reflection coefficient $R_C=|R_1|^2$ is
\ \\
\begin{equation}
R_C=\frac{(1+|\kappa|^2)^2(1-e^{-2|q_z| a})^2}{[(1+|\kappa|^2)^2(1-e^{-2|q_z| a})^2+16|\kappa|^2e^{-2|q_z| a}]}\;\;.
\end{equation}
Here again $T_C + R_C$ = 1, i. e. the current is conserved.  
\ \\

\subsection{TRANSMISSION FOR DIFFERENT ENERGY GAPS Pb$_{1-x}$Sn$_x$Se}
\ \\

Finally, in Fig. 5 we show oscillatory transmission coefficient $T_C$ versus barrier voltage calculated for three different Pb$_{1-x}$Sn$_x$Se compounds \cite{KD2016} characterised by decreasing energy gaps $E_g$ = 150 meV (for $x$=0), 20 meV (for $x$=0.16) and 2 meV (for $x$=0.18). The barrier width is $a=3$ nm and the electron energy $E$ = 5 meV above the bottom of conduction band in all cases. The effective mass $m^*_0$ is adjusted for each case according to Eq. (3). The resulting oscillatory patterns for all cases are similar to those shown in Figs. 2 and 3 but values of $T_C (V_b)$  strongly change for the decreasing gaps and the minimal value of $T_C$ increases for decreasing value of $E_g$.
\begin{figure}[b]
	\includegraphics[width=1.1\columnwidth, trim=0.3cm -1.cm 0.cm 0.cm]{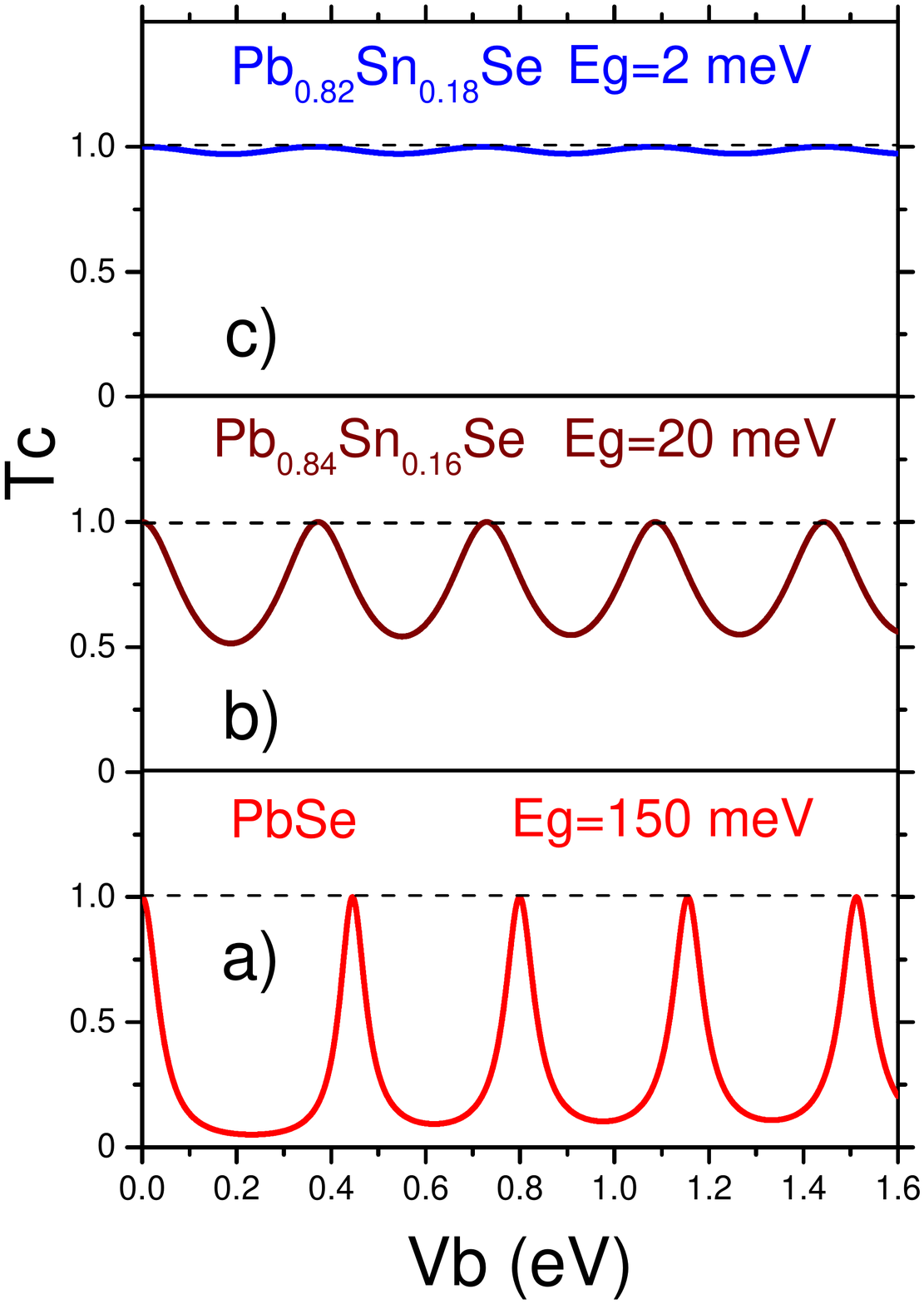}
	\caption{Oscillatory transmission probabilities $T_C$ versus applied voltage $V_b$ calculated for Pb$_{1-x}$Sn$_x$Se barrier potential of the width $a=3$ nm and for the forbidden gaps $E_g$ = 150 meV, 20 meV and 2 meV with value of $m^*_0$ adjusted (see Eq. (3)), respectively. Incoming electron energy is $E$ = 5 meV above the bottom of conduction band in each case.  For smaller value of $E_g$ $\kappa$ approaches 1 (see Eq. (14)) and consequently the minimum value of $T_C$ increases (see Eq. (22)). For $E_g$ = 0 there is $\kappa$ = 1 and hence  $R_1$ = 0, $R_2$ = 0, $T_1$ = 1 and $T_C$ = 1 for any value of $V_b$ (see Eqs. (15-17, 20)).}
	\label{fig5th}
\end{figure}

\section{DISCUSSION}

\ \\
In the following we briefly discuss physics underlying the results shown in the above figures.
As long as the barrier potential vanishes the incoming electrons do not experience any obstacle and the transmission $T_C$ is 1, as seen at the bottom of Fig. 2. However, when $V_b$ increases the electrons hit the forbidden gap of the barrier and $T_C$ falls quickly to zero, as seen on the left side of Fig. 2. For further growth of $V_b$ electrons are in the valence band of the barrier and values of $T_C$ increase. Thus the vanishing values of $T_C$ in Fig. 2 correspond roughly to the width $E_g$ of the forbidden gap. For the narrower barrier $a$ = 3 nm the damping of transmission is weaker and $T_C$ is always above zero (Fig. 3). At higher electron energy $E$ (see Fig. 4) the incoming electrons are damped in the forbiden gap at higher values of $V_b$: from $(E - E_g/2)$ to $(E + E_g/2)$. The purpose of Fig. 5 is to show that the phase and absolute values of $T_C(V_b)$ oscillations strongly depend on the forbidden energy gap of a semiconducting material. 
Comparing our results with these of Ref. \cite{Kats2006} dealing with tunneling in graphene we emphasize that the system we consider is more flexible than that of graphene just because of the possibility of changing the gap, whereas in monolayer graphene the gap is always zero. The second feature is of a more fundamental nature. It is emphasized in Ref. \cite{Kats2006} that, because in monolayer graphene one deals with chiral electron tunneling the transmission probability $T_C \equiv 1$ for any $V_b$. However, it is seen in the highest panel c) of our Fig. 5 that for the vanishing forbidden gap $E_g = 0$ in Pb$_{0.81}$Sn$_{0.19}$Se we would also have $T_C \equiv 1$  for any $V_b$ as a result of nonchiral electron tunneling. This coincidence does not seem fortuitous and it deserves further considerations.

\section*{Acknowledgments}
	The research was partially supported by
	the Foundation for Polish Science through the IRA Programme co-financed by EU within SG OP and the National Science Centre (Poland) through OPUS (UMO-2017/27/B/ST3/02470) project.
	
\ \\*[1cm]

\ \\	
\ \\

\ \\

\end{document}